\documentclass[11pt,a4paper]{article}

\usepackage[english]{babel}
\usepackage{latexsym}
\usepackage[latin1]{inputenc}
\usepackage{amsfonts} 
\usepackage{amssymb}
\usepackage{amsmath}
\usepackage{amsthm}
\usepackage{enumerate}
\usepackage{exscale}
\usepackage{epsfig}
\usepackage{fullpage}
\usepackage{textcomp}
\usepackage{microtype}
\usepackage{nicefrac}

\newtheorem{lemma}{Lemma}[section]
\newtheorem{theorem}{Theorem}

\title{\bf Randomized Approximation Schemes for the Tutte Polynomial and Random 
Clustering in Subdense and Superdense Graphs}
\author{Mathias Hauptmann\thanks{EFG Euskirchen.
    e-mail:{ \tt mathias.hauptmann@efg.euskirchen.de}}\:\:\:\:\:\quad\quad
   Ronja Tiling\thanks{EFG Euskirchen.}} 
\date{}

\begin{document}
\maketitle
\begin{abstract}
Extending the work of  \cite{AFW}, we show that there are randomized polynomial time approximation schemes for 
computing the Tutte polynomial in subdense graphs with an minimal node degree of
$\Omega\left ( \frac{n}{\sqrt{\log n}}\right )$ .
The same holds for the partition function $Z$ in the random cluster model with uniform edge probabilities and for the 
associated distribution $\lambda (A),\: A \subseteq E$ whenever the underlying graph $G=(V,E)$ is $c\cdot\frac{n}{\sqrt{\log (n)}}$-subdense. In the superdense case with node degrees $n-o(n)$, we show that the Tutte polynomial $T_G(x,y)$ is asymptotically equal to $Q=(x-1)(y-1)$. Moreover, we briefly discuss the problem of approximating $Z$ in the case of $(\alpha, \beta )$-power law graphs.



\end{abstract}

\section{Introduction}
Given a graph $G$, the Tutte polynomial $T_G(x,y)$ entails a tremendous amount of structural information about $G$. At $(x,y)=(1,1)$ it counts the number of spanning trees of $G$. It essentially contains the chromatic polynomial of $G$ via
$P(G,\lambda)=(-1)^{r(E)}\cdot\lambda^{k(G)}\cdot T_G(1-\lambda, 0)$,
where $k(G)$ denotes the number of connected components of $G$ and in general, for a subset
$A\subseteq E$ of edges of $G$, $r(A)=| V(G)|-k(A)$ with $k(A)$ being the number of connected components of the graph $(V(G),A)$.
A more comprehensive list of variants of the Tutte polynomial and its various specializations to hyperbolae
$H_{\alpha}=\{(x,y)|\: (x-1)(y-1)=\alpha\}$ can be found in \cite{W} and in \cite{AFW}.
As an immediate consequence, the Tutte polynomial is NP-hard to compute exactly, and also NP-hard to approximate for general graphs $G$.

It is therefore natural to investigate the approximability of $T_G(x,y)$ for special classes of graphs and also more special values $(x,y)$.
Alon, Frieze and Welsh give an RPTAS for the Tutte polynomial in everywhere dense graphs:
\begin{theorem}\label{afw_theorem}\cite{AFW}
For every $\epsilon >0$, there is an RPTAS for computing the Tutte polynomial
$T_G(x,y)$ in $\epsilon$-everywhere dense graphs with a minimum node degree of at least $\epsilon\cdot n$.  
\end{theorem}
The result relies on the general sampling method. 
In particular, it is shown there that the Tutte polynomial can be written as
\begin{eqnarray*}
  T_G(x,y) & = & \frac{y^m}{(x-1)(y-1)^n}\sum_{A\subseteq E}\left (\frac{y-1}{y}\right )^{|A|}\left (\frac{1}{y} \right )^{m-|A|}((x-1)(y-1))^{\kappa (A)}\\
               & = & \frac{y^m}{(x-1)(y-1)^n}\sum_{A\subseteq E} Q^{\kappa (A)}P(G_p=G_A)
\end{eqnarray*}
Thus $T_G(x,y)$ equals the expectation $E(Q^{\kappa (G_p)})$ with respect to the uniform edge probabilities $p=\frac{y-1}{y}$, where $Q=(x-1)(y-1)$.
Following the notion used in \cite{AFW}, we let $\zeta= y^m/((x-1)(y-1)^n)$. 

Concerning approximation lower bounds, in \cite{AFW} it is also shown that unless $NP=RP$, even in the dense case the 
Tutte polynomial cannot be computed exactly in polynomial time. 
\subsection{Our Results}
Our first result is an extension of the theorem \ref{afw_theorem} to the case of subdense graphs with a minimum node degree of order $n\slash\sqrt{\log (n)}$. 
\begin{theorem}\label{rm_tutte_theorem}
For every $\epsilon >0$, there is an RPTAS for the Tutte polynomial $T_G(x,y)$ in 
subdense graphs where for every vertex $v$, the node degree of $v$ satisfies
$d(v)\geq\epsilon\cdot\frac{n}{\sqrt{\log (n)}}$.
\end{theorem}
The method also applies to the general Random Cluster problem introduced by Fortuin and Kasteleyn \cite{FK}. 
\begin{theorem}
There is an RPTAS for computing the quantity 
$$Z=Z\cdot\sum_{A\subseteq E}\mu (A)\: 
=\:\sum_{A\subseteq E}p^{|A|}\cdot(1-p)^{|E\setminus A|}\cdot Q^{k(A)}$$
in the general Random Cluster bond percolation model on the edge set of a graph $G$
in the special case when edge probabilities are uniform ($p=p(e)$ for all $e\in E$) and the minimum node degree in $G$ is at least 
$\epsilon\cdot\frac{n}{\sqrt{\log (n)}}$.
\end{theorem}
Call a graph $G$ superdense if $d(v)=n-o(n)$ for all $v\in V$, where $n$ is the number of nodes of $G$. In particular, for a function
$f\colon {\mathbb N}\to {\mathbb Q}_+$ with $f(n)=o(n)$, graph $G$ is called $f(n)$-subdense if for all vertices $v\in V$,
$d(v)\geq n-f(n)$.
\begin{theorem}
For every $f(n)=o(n)$, the Tutte polynomial $T_G(x,y)$ of $f(n)$-superdense graphs asymptotically converges to $Q=(x-1)(y-1)$.
\end{theorem}
Finally we consider two variants of the Random Graph Model for power law graphs. For the case of $\beta\to\infty$, we indicate how to obtain estimates of the partition function $Z$ in this special case.
\section{Outline of the Sampling Method}
For the sake of completeness, we give a very brief outline of the general sampling method. See for instance the book by Mark Jerrum \cite{J} 
for a comprehensive treatement of the subject.
Suppose $X$ is a random variable. We consider the problem of computing the expectation 
$\mu=E(X)$. In case when this is an NP-hard computational problem, it might be convenient to approximately compute  $\mu$ by making use of sampling.
Let us assume that $X\colon\Omega\to \mathbb{Q}$ is such that we can sample efficiently, in the following sense: There is a polynomial time probabilistic algorithm $S$ which generates outputs independently at random and has the same distributiion as $X$. In particular, $E(S)=\mu$. The algorithm $S$ is called the \emph{sampler}. Consider then the following algorithmic approach:

\begin{enumerate}
\item Run the sampler $S$ $t$ times independently. Let $S_1,\ldots, S_t$ denote the associated outcomes.
\item Return $Z=\frac{1}{t}\cdot\left (S_1+\ldots + S_t \right )$
\end{enumerate}
Since $E(Z)=\mu$, using Chebyshev's inequality we obtain
$$Pr\left\{|Z-\mu |\:\geq\: \epsilon\right\}\quad\leq\quad \frac{Var (Z)}{\epsilon^2}$$
Since the $Z_i$ are independent, 
$$Var (Z)=\frac{1}{t^2}\cdot\sum_{i=1}^tVar(Z_i)=\frac{1}{t}\cdot Var (S)$$
Thus we obtain 
$$Pr\left\{|Z-\mu |\:\geq\: \epsilon\right\}\quad\leq\quad \frac{Var (Z)}{\epsilon^2}
\leq  \frac{Var(S)}{t\cdot \epsilon^2}$$
In case when $Var(S)$ is polynomially bounded in the size of the underlying graph, this yields
a polynomial time randomized approximation scheme. Suppose that $Var(S)\leq p(n)$ for some polynomial $p(n)$. Here $n$ denotes the input size. In the graph case, $n$ will be the number of vertices of the underlying graph $G$. For the sake of simplicity, we formulate the RPTAS in terms of an underlying graph $G$
such that $S$ is an associated sampler in the above sense.
\begin{itemize}
\item[] {\bf Sampling RPTAS for $\mu = E(S)$}\\
           Input: $G,\: \epsilon$\\
           Let $n$ be the number of vertices of $G$.\\
           Let $t$ be chosen such that $\frac{p(n)}{t\cdot\epsilon^2}\leq\frac{1}{2}$, i.e.
           $t=\left\lceil \frac{2\cdot p(n)}{\epsilon^2}\right\rceil$.\\
           Run $S$ $t$ times independently. Let $Z_1,\ldots , Z_t$ denote the associated outcomes.\\
           Return $\frac{1}{t}\cdot\left ( Z_1+\ldots +Z_t\right )$.
\end{itemize}
Note that this approach even yields an FPRAS for $\mu$.

\emph{Application to the Graph Case.}
In the case of the Tutte polynomial, the quantity the mean of which has to be estimated is 
$X=Q^{k(A)}$, where $A\subseteq E$ is the set of edges which are realized (i.e. open) within 
the underlying random model $G_p$. Following the lines of the proof of Thm 1 in \cite {AFW}, let $T$ denote the Tutte
polynomial and $Z=\zeta\cdot\frac{Z_1+\ldots + Z_t}{t}$ the output of the algorithm EVAL, where
$Z_i=Q^{\kappa (G_p)}$ is the result of the i-th sample $(i=1,\ldots t)$.
Compared to the above description of the general approach, the only additional step in the analysis is making use
of the inequality
$$Var(Z)=Var\left (\zeta\cdot\frac{Z_1+\ldots + Z_t}{t} \right )\:\leq\: \zeta^2\cdot Var (Z_i)\:\leq\: \zeta^2\cdot E(Z_i^2)$$
Whenever a polynomial bound for the expectation of $Z_i^2$ can be shown, this immediately yields an fpras for the above problem.

\section{An RPTAS for the Tutte Polynomial in Subdense Graphs}
In this section we give an RPTAS for the Tutte polynomial in $c\cdot\frac{n}{\sqrt{\log (n)}}$-subdense graphs. In order to do so, we follow the lines of the proofs in \cite{AFW}, adjusting them to the subdense case by carefully choosing the associated parameters. In Lemma \ref{components_G*} we construct an auxiliary graph $G^*$ which will then be used to obtain a bound on the mean of the quantity $Q^{2\kappa (G_p)}$.
$G^*$ is defined on the same vertex set as $G$ and $G_p$. 
The general idea is as follows: Vertices are adjacent in $G^*$ iff they are connected by sufficiently many 2-paths in $G$ and therefore are very likely to be in the same connected component in $G_p$.  \\
In the construction of $G^*$ we use two parameters $d_0$ and $d_1$, determining the edges and the number of components of this graph. 
Starting from the $ c \cdot \frac{n}{\sqrt {\log (n)}}$-subdense input graph $G$, the auxiliary graph $G^*=(V,E^*)$ is defined as follows: For any pair of vertices $u, v \in V$, $u v \in E^*$ if and only if $|N(u) \cap  N(v)|\geq d_0 \cdot \frac{n}{(\log (n))}$.
\begin{lemma}\label{components_G*}
The graph $G^*$ has at most $s = \frac{5}{2c} \cdot \sqrt{\log (n)}$ components.
\end{lemma}
\proof
(cf. Lemma 2 in \cite{AFW}) 
Suppose that the graph is defined in terms of parameter $d_0$ as above and 
that $G^*$ has more than $s=d_1 \cdot \log (n)$ components. 
We are now going to figure out for which choice of $d_0$ and $d_1$ this yields a contradiction. By assumption there exist vertices $v_1, v_2, ..., v_{s+1}$ which are in pairwise different components of $G^*$ and in particular non-adjacent to each other. This yields
\begin{eqnarray*}
  \left |\bigcup\limits_{i=1}^{s+1}N\left (v_i\right ) \right |
  & \geq &  \sum_{i=1}^{s+1} \left |N \left (v_i \right ) \right |
               - \sum_{i \neq j} \left |N \left ( v_i \right ) \cap N \left ( v_j \right ) \right | \\
  & > & (d_1 \cdot \sqrt{\log (n)}+1)\cdot c \cdot \frac{n}{\sqrt{\log (n)}}\}\\
  &    & \quad        -\frac{(d_1\cdot\sqrt{\log (n)})\cdot (d_1\cdot\sqrt{\log (n)} -1)}{2}\cdot d_0
          \cdot \frac{n}{(\log (n))} \\
  & \geq & d_1 \cdot c \cdot n - {d_1}^2 \cdot d_0 \cdot n \\
  & = & (-{d_1}^2 \cdot d_0 + d_1 \cdot c)\cdot n
\end{eqnarray*}
We now want to determine $d_0$ and $d_1$ such that this bound becomes $>n$, which yields the desired contradiction. Since the term $R(d_1)=-{d_1}^2 \cdot d_0 + d_1 \cdot c-1$ is quadratic in $d_1$, we choose $d_1=\frac{c}{2d_0}$. Then in order to obtain 
$$R(d_1)=R\left (\frac{c}{2d_0} \right )
= \frac{1}{d_0}\cdot\left (-\frac{c^2}{4}+\frac{c^2}{2} \right )-1\cdot>0,$$
we choose $d_0=\frac{c^2}{5}$. This concludes the proof of the lemma. \qed

\begin{lemma}\label{expectation}
Let $s=O(\sqrt{\log n})$ be the upper bound for the number of components of $G^{*}$ in the 
$\frac{n}{\sqrt{\log (n)}}$-subdense case. For $Q\geq 1$, 
$$E\left ( Q^{2\kappa (G_p)}\right )\:\:\leq\:\: 2Q^{2s}\:\: =\:\: n^{O(1)}.$$
\end{lemma}
\proof In the proof we follow the lines of the proof of Lemma 3 in [AFW], adjusting it to 
the subdense case. In particular, the expectation of the random variable 
$Q^{2\kappa (G_p)}$ is upper bounded by a sum of estimates of certain interval probabilities, which will turn out to yield an upper bound which is a power series.

By definition, 
$$E\left ( Q^{2\kappa (G_p)}\right )=\sum_{k=1}^{n}Q^{2k}\cdot P(\kappa (G_p)=k)$$
In order to obtain appropriate bounds for the mean, consider the following events 
$$\mathcal{E}_u=\{\kappa (G_p)> u\cdot\log (n)\},\quad u=1,\ldots , \frac{n}{\log (n)},$$
and let $\rho$ denotes the precise number of components of $G^*$. In the preceeding lemma we have shown that
$\rho\leq s=\nicefrac{5}{2c}\cdot\sqrt{\log (n)}$, where $c$ is the subdensity parameter.
The probability of event $\mathcal{E}_u$ can be bounded as follows.
\begin{eqnarray*}
  P(\mathcal{E}_u) & \leq &
        \begin{array}{l}
           \mbox{}\\
           P\left (\exists C_i\:\mbox{connected component of $G^*$ containing}\right.\\
           \quad\quad\left.\mbox{vertices from $u\sqrt{\log (n)}$ distinct components of $G_p$}\right ) 
          \end{array}\\
  & \leq & \begin{array}{l}
           \mbox{}\\
           P\left (\exists C_i\:\mbox{connected component of $G^*$ containing}\right.\\
           \quad\quad\left.\mbox{vertices from $u+1$ distinct components of $G_p$}\right ) 
          \end{array}\\
  & \leq & \begin{array}{l}
                 \mbox{}\\ 
                 P\left (\exists C_i\:\mbox{connected component of $G^*$ containing}\right.\\
                 \quad\quad \mbox{vertices $x_1,\ldots , x_{u+1}$ such that for $i=1,\ldots ,\frac{u}{2}$, }\\
                 \quad\quad \left.\mbox{$x_{2i-1}x_{2i}$ is an edge in $G^*$ but not in $G_p$}\right )
               \end{array}\\
  & \leq & {n\choose u+1}\cdot (1-p^2)^K, \quad \mbox{where $K=\left (d_0\cdot\frac{n}{\log (n)}-2u\right )\cdot\frac{u}{2}$}\\
  &       &  \quad\quad \left (\mbox{note that the term $-2u$ is due to the fact that for $x_{2i-1}, x_{2i}$, some of the}\right.\\
  &       &  \quad\quad \mbox{2-paths connecting these two nodes might contain other nodes $x_j$}\\ 
  &       &  \quad\quad \left.\mbox{and thus some edges might be counted twice}\right )\\
  & \leq & n^{u+1}\cdot e^{\log (1-p^2)\cdot (\frac{c^2}{5}\cdot\frac{n}{\log (n)}-2u)\cdot\frac{u}{2}}\\
  & =    &  e^{(u+1)\log (n) -\log (\frac{1}{1-p^2})\cdot  (\frac{c^2}{5}\cdot\frac{n}{\log (n)}-2u)\cdot\frac{u}{2}}\\
  & \leq &  \left ( e^{2\log (n)-\log (\frac{1}{1-p^2})\cdot  (\frac{c^2}{5}\cdot\frac{n}{\log (n)}-2u)\cdot\frac{1}{2} }\right )^{u}\\
  & \leq &  \left ( e^{-C\cdot \frac{n}{\log (n)}}\right )^{u }\quad\mbox{for some constant $C>0$ depending only on $p, c$}\\
  &       &   \quad\quad\mbox{(for $c$ being sufficiently large - note that $c>1$ is possible for sublinear densities)}
\end{eqnarray*}
We obtain
\begin{eqnarray*}
E(Q^{2\kappa (G_p)}) & \leq & Q^{2\rho}\cdot P(1\leq \kappa (G_p)\leq \rho)\:
                                              +\sum_{u=1}^{\frac{n}{\log (n)}}Q^{2(u+1)}\cdot P(\mathcal{E}_u)\\
                                  & \leq & Q^{2\rho}\cdot 1 \: +\sum_{u=1}^{\frac{n}{\log (n)}}Q^{2(u+1)}\cdot \left ( e^{-C\cdot \frac{n}{\log (n)}}\right )^{u }\\
                                  & \leq & Q^{2\rho}\: +\: Q^2\cdot\sum_{u=1}^{\frac{n}{\log (n)}}\left ( e^{4\log Q-C\cdot\frac{n}{\log (n)}}\right )^{u}
\end{eqnarray*}
For $q(n)=e^{4\log Q-C\cdot\frac{n}{\log (n)}}$, this last bound is equal to 
$$Q^{2\rho}+Q^2\cdot q\cdot \frac{1-q(n)^{n\slash\log (n) +1}}{1-q(n)}= Q^{2\rho}\:\: +o(1),$$
since $q(n)\longrightarrow 0$ as $n$ tends to infinity. This concludes the proof of the lemma. \qed

\begin{theorem}
Let $G$ be a $c\cdot\frac{n}{\sqrt{\log (n)}}$-subdense graph, $T=T_G(x,y)$ its Tutte polynomial
and $Z$ the output of the sampling algorithm with 
$Z=\frac{\zeta}{t}\cdot\left (Z_1+\ldots + Z_t\right )$.
Then 
$$P\left (|Z-T|\:\geq\: \epsilon\cdot T\right )\:\leq\: \frac{1}{4}$$
\end{theorem}
\proof
We restrict ourselves to the case $Q\geq 1$, where $Q=(x-1)(y-1)$. The other case 
s proved in a similar manner.
Using Chebyshev's inequality and the fact that $T=E(Z)$, 
$$P\left (|Z-T|\geq\epsilon\cdot T\right )\:\leq\: \frac{Var(Z)}{\epsilon^2\cdot T^2}\:\leq\: 
\frac{\zeta^2}{\epsilon^2\cdot t}\cdot\frac{Var(Z_i)}{T^2}\:\leq\: 
\frac{\zeta^2}{\epsilon^2\cdot t}\cdot\frac{E(Z_i^2)}{T^2}$$
\qed
\section{Approximating the Distribution Function of the Random Cluster Model}
In the \emph{Random Cluster Model} introduced by Fortuin and Kasteleyn \cite{FK}, 
we are given a graph $G=(V,E)$ and edge probabilities $p_e\in [0,1]\: (e\in E)$. The associated probability distribution
on the power set of $E$ is given by
$$\mu (A)=Z^{-1}\cdot\sum_{A\subseteq E}\left (\prod_{e\in A}p_e\right )\:\cdot\: 
\left (\prod_{e\notin A}(1-p_e)\right )\cdot Q^{\kappa (A)},$$
where $Z$ is defined in such a way that $\sum_{A\subseteq E}\mu (A)\: =\: 1$. $Z$ is called the \emph{partition function} of the model.

The probability that a particular set of edges is open - in the sense of being among the edges which are 
realized in the current sample - yields the so called \emph{distribution function} $\lambda$, given by
$$\lambda (A)\: =\: \sum_{X\colon\: X\supseteq A}\mu (X)$$
We show in this section that the sampling method also yields an FPRAS for the partition function $Z$ as well as the distribution function $\lambda$
in the cases of dense and of  $c\cdot\frac{n}{\sqrt{\log (n)}}$-subdense graphs.
\begin{theorem}
For the Random Cluster Model in $c\cdot\frac{n}{\sqrt{\log (n)}}$-subdense graphs with $c$ being sufficiently large, the following holds.
\begin{itemize}
\item[(a)] There is an FPRAS for the partition function $Z=\sum_{A\subseteq E}p^{|A|}\cdot (1-p)^{|E\setminus A |}\cdot Q^{\kappa (A)}$.
\item[(b)] There is an FPRAS for the distribution function $\lambda (A)$. In particular, this algorithm gets as an input an underlying 
               graph $G$, the value $Q$ as well as a subset $A$ of the edges of $G$ and some approximation parameter $\epsilon$
               and returns as an output an $\epsilon$-approximation of the value $\lambda (A)$. 
\end{itemize} 
\end{theorem}
\proof
Concerning (a), we just observe that approximating $Z$ is essentially the same problem as estimating the Tutte polynomial 
$T_G(x,y)$, except that we do not have $p$ and $Q$ depending on $x$ and $y$. Since the FPRAS for the Tutte polynomial does 
not rely on the special choice of $p$ and $Q$, precisely the same algorithm and proofs work for the case of general $Z$.

Concerning (b), we observe that 
\begin{eqnarray*}
\lambda (A) & = & Z^{-1}\cdot \sum_{X\supseteq A}p^{|X|}\cdot (1-p)^{|E\setminus A|}\cdot Q^{\kappa (X)}\\
                  & = & Z^{-1}\cdot p^{|A|}\cdot\sum_{X\supseteq A}p^{|X\setminus A |}\cdot (1-p)^{|E\setminus X|}
                           \cdot 1^{|A|}\cdot Q^{\kappa (A)}\\
\end{eqnarray*}
which is simply a product probability distribution term wrto the edge probabilities 
$$q(e)\quad = \quad \left\{
                                  \begin{array}{ll}
                                      p, & e\not\in A\\
                                      1, & e\in A
                                  \end{array}
                               \right.$$
We observe that $\lambda (A)$ is equal to 
the product of  the quantity $Z^{-1}$ in the graph $G$ and the quantity $Z_{G\slash A}$ in the graph $G\slash A$ obtained from $G$ by contracting the set of edges $A$, both with edge probabilities $p$:
\begin{eqnarray*}
  \lambda (A)  & = & Z^{-1}\sum_{X\subseteq V(G\slash A)}p^{|X|}\cdot (1-p)^{|E(G\slash A)\setminus X|}\cdot Q^{\kappa (X)}\:\:
                              =\:\: Z^{-1}\cdot Z_{G\slash A}
\end{eqnarray*} 
where in that case $\kappa (X)$ is the set of components of the graph $(V(G\slash A), X)$.
From our previous results it follows that there are fully polynomial time approximation schemes for both $Z$ and $Z_{G\slash A}$.
Now the statement follows immediately from the general fact that 
if $A$ and $B\neq 0$ are computational problems which both provide an fpras, then there is an fpras for $A\slash B$.
In order to prove this, 
suppose $T_A$ and $T_B$ are fully polynomial time approximation schemes for $A$ and $B$ respectively. It suffices to show that if $T_A(x)$ and $T_B(y)$ are
$(1+\epsilon)$-approximate solutions to $A(x)$ and $B(y)$ respectively, then  $T_A(x)\slash T_B(y)$ is a $(1+\epsilon )$-approximate solution for $A(x)\slash B(y)$.
We have by assumption that $(1+\epsilon )^{-1}A(x)\leq T_A(x)\leq (1+\epsilon )A(x)$ and 
$(1+\epsilon )^{-1}B(x)\leq T_B(x)\leq (1+\epsilon )B(x)$. This yields 
$$\frac{T_A(x)}{T_B(y)}\leq \frac{1+\epsilon}{(1+\epsilon)^{-1}}\cdot \frac{A(x)}{B(x)}=(1+\epsilon)^2\cdot \frac{A(x)}{B(x)}$$
as well as 
$$\frac{T_A(x)}{T_B(x)}\geq \frac{1}{(1+\epsilon)^2}\frac{A(x)}{B(x)},$$
which concludes the proof ot the theorem.
\qed

For superdense graphs, we obtain the 
following result.
\begin{theorem}
For every $f(n)=o(n)$, 
$$E\left (Q^{\kappa (G_p)}\right )\:\longrightarrow\: Q\quad \mbox{as $n\to\infty$}$$
and therefore
$$T_G(x,y)\:\longrightarrow\: \frac{y^m}{(y-1)^n}\quad (n\to\infty).$$
\end{theorem}
\proof
We give a direct estimate of $E(Q^{\kappa (G_p)})$ as follows: 
\begin{eqnarray*}
E\left (Q^{\kappa (G_p)} \right ) & \leq & \sum_{i=1}^nP(\kappa (G_p)\geq i)\cdot Q^{i}\\
 & \leq & \sum_{i=1}^{n}P(\exists u_1,\ldots , u_i\:\mbox{pairwise not adjacent})\cdot Q^{i}\\
 & \leq & \sum_{i=2}^n n^{i}\cdot (1-p)^{\frac{i(i-1)}{2}\cdot (n-2f(n))}\cdot Q^{i}
              \quad + P(G_p\:\mbox{connected})\cdot Q\\
 & \leq & \sum_{i=2}^{n} n^{i}\cdot (1-p)^{i\cdot (n-2f(n))}\cdot Q^{i}\quad + P(G_p\:\mbox{connected})\cdot Q\\
 & =    & \sum_{i=2}^{n} (\underbrace{n\cdot (1-p)^{n-2f(n)}\cdot Q }_{=:\: r=r_{n,p}})^{i}
              \quad + P(G_p\:\mbox{connected})\cdot Q\\
 &       & \quad\mbox{(note that the term $r$ only depends on $n, Q$ and $p$ but not on the index $i$)}\\
 & =    & r^2\cdot\sum_{i=0}^{n-2} r^{i}\quad\quad + P(G_p\:\mbox{connected})\cdot Q\\
 & =    & r^2\cdot \frac{1-r^{n-1}}{1-r}\quad\quad +P(G_p\:\mbox{connected})\cdot Q
\end{eqnarray*}
 Since $Q$ is constant for given $x$ and $y$, $r\longrightarrow 0\: (n\to \infty)$. Moreover,
$$P(G_p\:\mbox{connected})=1-P(\exists\: u\neq v\: (\mbox{$u$ not connected to $v$}))\geq 1-n^2\cdot (1-p)^{n-2f(n)}$$
This implies that $P(G_p\:\mbox{connected})\to 1\: (n\to\infty)$ and therefore 
$E(Q^{\kappa (G_p)})\leq Q (1+o(1))$. It remains to show that $E(Q^{\kappa (G_p)})\geq Q (1-o(1))$.
We have
\begin{eqnarray*}
E(Q^{\kappa (G_p)}) & = & \sum_{i=1}^{n}P(\kappa (G_p)=i)\cdot Q^{i}\\
                                & \geq & P(\kappa (G_p)=1)\cdot Q\\
                                & =     & (1-P(\kappa (G_p) \geq 2)\cdot Q\\
                                & \geq & \left (1-\sum_{u\neq v, \: u,v\in V }P(\mbox{$u, v$ not adjacent in $G_p$})\right )\cdot Q\\
                                & =     &  (1-\underbrace{n^2\cdot (1-p)^{n-2f(n)}}_{\to 0\: (n\to\infty)}  )\cdot Q\\
                                & \geq & (1-o(1))\cdot Q
\end{eqnarray*}
which completes the proof of the theorem.
\qed

\section{The Tutte Polynomial of Power Law Graphs}
A graph or multigraph whose node degree distribution follows a power law is called a \emph{Power Law Graph}. This means that 
the number of nodes of degree $i$ is approximately proportional to $i^{-\beta}$, where $\beta$ is the negative slope of 
the degree sequence on a log-log scale, i.e. the log-log growth rate of the graph.
We consider here the model of Aiello, Chung and Lu \cite{ACL}. According to this model, 
a graph or multigraph $G$ is called an $(\alpha,\beta )$-Power Law Graph, denoted as 
$(\alpha, \beta )$-PLG, if the maximum node degree of $G$ is 
$\Delta = \left\lfloor e^{\alpha\slash \beta}\right\rfloor$ and moreover, for $i=1,\ldots \Delta$, the number of nodes 
of degree $i$ ist equal to $\left\lfloor \frac{e^{\alpha}}{i^{\beta}}\right\rfloor$.
It has been shown in \cite{ACL} that the number of vertices $n$ and the number of edges $m$ of $(\alpha, \beta)$-PLGs asymptotically behave as follows:
$$n=\left\{\begin{array}{ll}
  \zeta (\beta)\cdot e^{\alpha}, & \beta >1\\
  \alpha\cdot e^{\alpha}, & \beta =1\\
  \frac{e^{\alpha\slash\beta}}{1-\beta}, & 0<\beta <1
\end{array}\right.\quad\quad
m=\left\{\begin{array}{ll}
  \frac{1}{2}\cdot \zeta (\beta -1)\cdot e^{\alpha}, & \beta >2\\
  \frac{1}{4}\cdot \alpha\cdot e^{\alpha}, & \beta =2\\
  \frac{1}{2}\cdot\frac{e^{2\alpha\slash\beta}}{2-\beta}, & 0<\beta <2
\end{array}\right.$$
A framework for random graphs with a given degree sequence has been investigated by Molloy and Reed \cite{MR}.
Aiello, Chung and Lu specialize this framework to the case of $(\alpha, \beta)$-PLGs. A random PLG is generated as follows. Given 
$\alpha$ and $\beta$, take $i$ copies of each node which is intended to have degree $i$. Let $L$ be the set of all these copies.
Then generate a perfect matching on $L$ uniformly at random. This induces an $(\alpha, \beta )$-power law multigraph.

Here we are concerned with the task of estimating or approximating the partition function $Z$ and the distribution function $\lambda$ of the 
Random Cluster Model in power law graphs.
In particular, we consider the following two variants. \\[1.7ex]
\emph{First Model: Sampling from a given PLG.} In the first model, we are given a connected $(\alpha,\beta )$-PLG as an input graph and edge probabilities 
$p=(p(e))$. Then we generate a subgraph $G_p$ of $G$ at random with respect to $p$. We want to determine the associated partition function $Z$.\\[1.7ex] 
\emph{Second Model: Generating a PLG uniformly at random.} In the second model we generate a power law graph at random, using the matching-based 
algorithm from \cite {ACL}. In that model, we want to compute or estimate $Z$ and $\lambda$ with respect to this distribution. \\[1.7ex]
Molloy and Reed \cite{MR} analyze the connectivity properties of random graphs with given degree sequences in terms of the 
quantity $Q=\sum_{i\geq 1}i(i-2)\lambda_i$, where $\lambda_i=\lim_{n\to\infty}d_i(n)\slash n$ and $d_i(n)$ is the number of nodes of degree $i$ in a graph 
with $n$ nodes in this model.

For the sake of completeness, we compute here $Q$ in case of a power law distribution (cf. also \cite{ACL}).\\[0.9ex]
{\sl Case $\beta >1$.}
We have 
\begin{eqnarray*}
Q & = & \lim_{\alpha\to\infty}\sum_{x=1}^{e^{\alpha\slash\beta}}x(x-2)\frac{\frac{e^{\alpha}}{x^{\beta}}}{\zeta (\beta )\cdot e^{\alpha}}
    = \lim_{\alpha\to\infty}\frac{1}{\zeta (\beta )}\sum_{x=1}^{e^{\alpha\slash\beta}}x^{2-\beta}-2x^{1-\beta}
    = \frac{\zeta (\beta -2)-2\zeta (\beta -1)}{\zeta (\beta)}
\end{eqnarray*}
Thus in the case $\beta >1$, the quantity $Q$ converges.\\[0.9ex]
{\sl Case $\beta <1$.} Then
\begin{eqnarray*}
 Q & = & \lim_{\alpha\to\infty}\sum_{x=1}^{e^{\alpha\slash\beta}}x(x-2)\frac{\frac{e^{\alpha}}{x^{\beta}}}
                                                                                                                  {\frac{e^{\alpha\slash\beta}}{1-\beta}}
       \quad =\quad  \lim_{\alpha\to\infty}\frac{e^{\alpha (1-\frac{1}{\beta})}}{1-\beta}
          \cdot\sum_{x=1}^{e^{\alpha\slash\beta}}\left (x^{2-\beta}-2x^{1-\beta} \right )\\
     & \approx & \lim_{\alpha\to\infty}\frac{e^{\alpha (1-\frac{1}{\beta})}}{1-\beta}\cdot
          \left [\frac{x^{3-\beta}}{3-\beta}-\frac{2x^{2-\beta}}{2-\beta}  \right ]_1^{e^{\alpha\slash\beta}}
    \quad = \quad \lim_{\alpha\to\infty}\frac{e^{\alpha\cdot (1-\frac{1}{\beta} )}}{1-\beta}
                   \cdot\left (\frac{e^{\alpha\cdot\frac{3-\beta}{\beta}}-1}{3-\beta}
                                 -\frac{2e^{\alpha\cdot\frac{2-\beta}{\beta}}-2}{2-\beta} \right )\\
     & = & \lim_{\alpha\to\infty}\frac{1}{(1-\beta )(3-\beta )}
                                      \cdot\frac{e^{\alpha\cdot\frac{3-\beta}{\beta}}-1}{e^{\alpha\cdot\frac{1-\beta}{\beta}}}
              -\frac{2}{(1-\beta)(2-\beta)}\cdot\frac{e^{\alpha\cdot\frac{2-\beta}{\beta}}-2}{e^{\alpha\cdot\frac{1-\beta}{\beta}}}\\
     &\approx & \lim_{\alpha\to\infty}\frac{1}{(1-\beta )(3-\beta )}
                         \cdot e^{\alpha\cdot\frac{3-\beta- (1-\beta)}{\beta}}   
                      - \frac{2}{(1-\beta)(2-\beta)}\cdot   e^{\alpha\cdot\frac{2-\beta- (1-\beta)}{\beta}}  \\
     & = &        \lim_{\alpha\to\infty}\frac{1}{(1-\beta )(3-\beta )}
                         \cdot e^{\nicefrac{2\alpha}{\beta}}  
                      - \frac{2}{(1-\beta)(2-\beta)}\cdot   e^{\nicefrac{\alpha}{\beta}}
\end{eqnarray*}
We see that the term diverges to infinity as $\alpha\to\infty$, so at least for every $\alpha$ the term is positive.\\[2.5ex]
We are now going to address the two models from above. Here we only consider one extreme case. A more extensive discussion will be provided in a subsequent extended version of this paper.

Observe that $\lim_{\beta\to\infty}\frac{1}{2}\zeta (\beta -1)\slash\zeta (\beta) =\frac{1}{2}$. This implies that the number of components of an $(\alpha, \beta)$-PLG
is asymptotically equal to $\frac{n}{2}$ when $\beta\to\infty$. Thus, if $G_{\alpha,\beta}$ denotes a graph or multigraph drawn at random from $PLG(\alpha,\beta)$, then  
in the second model, the partition function of the associated random cluster model is asymptotically equal to $Q^{n\slash 2}$. In the first model, basically we draw a random
subgraph $G_p$ from a graph $G$ that is a perfect matching on the set of nodes. The number of connected components is then within the interval 
$\left [ \frac{n}{2},\: n\right ]$, with a binomial distribution. Further details will be given in an extended version of this paper.
 $\mbox{}$\\
$\mbox{}$\\
{\bf Acknowledgement.} We would like to thank the executive board of the Emil-Fischer-Gymnasium Euskirchen, 
Prof. Dr. Michael Szczekalla and Dr. Wolfram Ferber, for their continuous encouragement and kindness. This work is supported by
the LemaS iniciative, TP Individuelle Lernpfade. In particular we want to express our gratitude to Prof. Dr. Heidrun St{\"o}ger, 
Prof. Dr. Albert Ziegler and their team, especially Kathrin Emmerdinger, Sonja Beyer and Tina-Myrica Daunicht.
Ronja would like to thank her parents and her family for guidance and support.
\newpage



\begin{thebibliography}{alpha}

\bibitem[ACL]{ACL}
 William Aiello, Fan Chung, Linyuan Lu,
 \textit{A random graph model for power law graphs},
 Experimental Mathematics 10(1), pp. 53--66, 2001.

\bibitem[AFW]{AFW}
 Noga Alon, Alan Frieze, Dominic Welsh,
 \textit{Polynomial time randomized approximation schemes for
           Tutte-Groethendieck invariants: the dense case},
 ECCC TR94-005, 1994.

\bibitem[FK]{FK}
 Fortuin, C.M and Kasteleyn, P.W.,
 \textit {On the random cluster model I. Introduction and relation to other models},
 Physica 57, pp. 536--564,
 1972.

\bibitem[J]{J}
  Mark Jerrum,
 \textit{Counting, Sampling and Integrating: Algorithms and Complexity},
 Lectures in Mathematics, ETH Z{\"u}rich, Birkh{\"a}user, 2003.
\bibitem[MR]{MR}
 Michael Molloy, Bruce Reed,
 \textit{A critical point for random graphs with a given degree sequence},
 Random Structures and Algorithms, Vol. 6, Issue 2-3, pp. 161--180, 1995.

\bibitem[OW]{OW}
 J.G. Oxley, D.J.A. Welsh,
 \textit{The Tutte polynomial and percolation},
 Graph Theory and Related Topics (eds. JA. Bondy, U.S.R. Murty), 
 Academic Press, London, pp. 329--339, 1979.

\bibitem[W]{W}
 Dominic Welsh,
 \textit{Complexity: Knots, Colourings and Counting},
 London Mathematical Society LNS 186, Cambridge University Press, 1993.


\end{thebibliography}
\end{document}